\documentclass[preprint, twocolumn]{aastex63}

\usepackage{microtype,longtable,verbatim,float,booktabs,graphicx,tabularx}
\usepackage{amsmath} 
\usepackage{textcomp,multirow,makecell}

\newcommand{\SFinf}{$\rm SF_{\infty}$}
\newcommand{\redchi}{$\chi_{\nu}^2  \;$}

\setcitestyle{aysep={}}
\received{\today}
\revised{---}
\accepted{---}

\submitjournal{ApJ}

\shorttitle{Rubin Survey Design Consequences for Binary SMBHs}
\shortauthors{Davis et al.}

\begin{document}

\title{The Consequences of Rubin Observatory Time-Domain Survey Design and Host-Galaxy \\ Contamination on the Identification of Binary Supermassive Black Holes}

\correspondingauthor{Megan C. Davis}
\email{megan.c.davis@uconn.edu}

\author[0000-0001-9776-9227]{Megan C. Davis}
\altaffiliation{NSF Graduate Research Fellow}
\affiliation{Department of Physics, 196A Auditorium Road, Unit 3046, University of Connecticut, Storrs, CT 06269, USA}

\author[0000-0002-1410-0470]{Jonathan R. Trump}
\affiliation{Department of Physics, 196A Auditorium Road, Unit 3046, University of Connecticut, Storrs, CT 06269, USA}

\author[0000-0003-3579-2522]{Maria Charisi}
\affiliation{Department of Physics and Astronomy, Washington State University, Pullman, WA 99163, USA}
\affiliation{Institute of Astrophysics, FORTH, GR-71110, Heraklion, Greece}

\author[0000-0001-8557-2822]{Jessie C. Runnoe}
\affiliation{Department of Physics and Astronomy, Vanderbilt University, Nashville, TN 37235, USA}
\affiliation{Department of Life and Physical Sciences, Fisk University, 1000 17th Avenue N, Nashville, TN, 37208, USA}

\author[0000-0002-0167-2453]{W. N. Brandt}
\affiliation{Department of Astronomy \& Astrophysics, 525 Davey Lab, The Pennsylvania State University, University Park, PA 16802, USA}
\affiliation{Institute for Gravitation and the Cosmos, The Pennsylvania State University, University Park, PA 16802, USA}
\affiliation{Department of Physics, 104 Davey Lab, The Pennsylvania State University, University Park, PA 16802, USA}




\author[0009-0001-6853-9470]{Kaylee E. Grace}
\affiliation{Department of Physics \& Astronomy, 217 Sharp Lab, Newark, DE 19716, USA}


\author[0009-0001-4123-3957]{London E. Willson}
\affiliation{Department of Physics, 196A Auditorium Road, Unit 3046, University of Connecticut, Storrs, CT 06269, USA}

\begin{abstract}
Binary supermassive black holes (SMBHs) are consequences of galaxy mergers and dominate the low-frequency gravitational wave background. Finding binary SMBHs in existing time-domain observations has proven difficult, as their periodic, electromagnetic signals can be confused with the natural variability of single quasars. 
In this work, we investigate the effects of host-galaxy contamination and survey design (cadence and duration) on the detectability of binary SMBHs with the upcoming Rubin Observatory Legacy Survey of Space and Time (LSST). We simulate millions of LSST light curves of single and binary quasars, with a distribution of quasar and host-galaxy properties motivated by empirical observations and the anticipated LSST detection space. We then apply simple sinusoidal curve fits as a potential, computationally inexpensive detection method. We find that host-galaxy contamination will increase false-positive rates and decrease binary parameter recovery rates. Lower mass, lower luminosity binary systems are most likely to be negatively affected by host galaxy contamination. We also find that monitoring duration affects binary detection more than survey effective cadence for this detection method. As the light curve duration increases, false-positive rates are suppressed and binary parameter recovery rates, especially for binary period, are improved. Increasing the light curve duration from 5 to 10 yrs shows the most dramatic  improvement for successful binary detection and false-positive rejection, with additional improvement from extending the light curve duration to 20 yrs. The observation duration increase is especially critical for recovering binary periods that are longer than a decade. 

\end{abstract}


\section{Introduction}\label{sec:intro}

Sub-parsec binary SMBHs will be the loudest gravitational wave (GW) sources in the Universe, constituting the low-frequency gravitational wave background \citep{Sesana_2008,Christensen_2018, Kelley2019c, Kelley2021, 2023nanograv15yearBSMBHConstraints, 2023Nanograv15yrEvidence, Antoniadis2023secondDR, Reardon2023, Milesetal2025MeerKAT}. Identifying binary SMBHs electromagnetically will reveal promising gravitational wave sources. The localization of targets makes a gravitational wave search more sensitive, making these sources easier to detect with pulsar timing arrays, like the North American Nanohertz Observatory for Gravitational Waves \citep[NANOGrav;][]{FirstNANOgravWhitePaper, NANOgrav_McLaughlin2013, NANOGrav2019}, and the upcoming space-based gravitational wave observatory, LISA \citep{LISA}. The binary SMBH population will also constrain galaxy merger rates, which are crucial to our understanding of both SMBH-galaxy co-evolution and the hierarchical structure formation of galaxies \citep[see][and references therein]{Merloni_2010, Comerford}.

There exist many confirmed detections of dual Active Galactic Nuclei (AGN), where the SMBH are not gravitationally bound to each other but are embedded within the same host galaxy \citep[e.g.,][]{VanWassenhove2012,De_Rosa_2019,DualAGN_2023}. Dual AGN are expected to be the early-evolution precursors to binary SMBH, where the nuclei are gravitationally bound and form a Keplerian orbit \citep{De_Rosa_2019}. 
We have hundreds of optical, photometric binary SMBH candidates \citep[i.e.,][]{Graham2015, Charisi2016, Liu2019,Chen2020,Chen2024}, with no full confirmation yet. 

The binary SMBH search in existing and ongoing time-domain observations of quasars is a search for sustained periodicity. The expected binary SMBH signals are described further in Section \ref{sec:datgen}. However, intrinsic quasar variability can appear quasi-periodic and a single quasar can masquerade as a periodic signature of a binary SMBH \citep{Vaughan2016, Davis_2024}. Furthermore, time-domain surveys that are limited in cadence and/or duration produce high binary SMBH false-positive rates \citep{Vaughan2016, Barth_2018, Zhu2020, Witt_2022, Robnik2024}. Current detection methods are also sensitive to quasar variability amplitudes and can report false-positive detections and incorrect recovered binary parameter values due to the intrinsic quasar variability present in binary systems \citep{Vaughan2016, Davis_2024}.  Constraints on 
the gravitational wave background indicate false-positive contamination of the binary quasar candidate population: the current number of candidates significantly exceeds the number of expected binaries from these constraints \citep{Zhu2014_PTAs,Sesana_2018}.

Upcoming massive time-domain surveys, like the Vera C. Rubin Observatory \citep{2019Rubin_Ivezic}, may assist with the binary SMBH detection challenge \citep{Witt_2022}.
First light for Rubin was in 2025\footnote{https://www.lsst.org/about/project-status}. Rubin will obtain photometric, $ugrizy$ coverage during its 10-yr Legacy Survey of Space and Time (LSST). LSST is expected to produce time-domain monitoring that has a cadence and duration appropriate to detect potential sub-parsec binary SMBHs \citep{Ivezic2017, Charisi_2022}. 
The main survey, Wide-Fast-Deep (WFD), will utilize 80-90\% of the total survey observation time and will observe nearly 18,000~deg$^2$ of the southern hemisphere sky. The remaining 10-20\% of survey time will be allocated to other observing programs, including the Deep Drilling Fields (DDF; 9.6~deg$^2$ each, though the EDF-S is larger by a factor of $\sim$2), which will receive a rapid effective cadence and deeper on-sky coverage \citep{Bianco_2022LSST}. We explore the binary detection potential for both the WFD and DDFs in this work.

LSST is expected to produce nearly 20 terabytes of data every night\footnote{https://www.lsst.org/about/dm}. LSST will also observe nearly 20 - 100 million AGN \citep{lsstsciencecollaboration2009lsst, XinHaiman2021}. 
Initial binary detection methods will need to be computationally inexpensive to triage false-positive single quasar light curves from current and future binary SMBH candidate lists. 
Our work highlights a computationally inexpensive initial detection method, sinusoidal curve fits, where we include strengths, weakness, and recommendations for Rubin applications. We also discuss survey design needs for initial binary SMBH triage.

In the next section, Section \ref{sec:datgen}, we review the construction and underlying models of the Rubin LSST light curves. Section \ref{sect:Analysis} details the simple sinusoidal curve fitting. We present the results of the light curve fitting, including predicted false positive rates and population statistics, in Section \ref{sec: results}. We discuss the results, with recommendations for future Rubin LSST observations and comparisons with gravitational wave searches, in Section \ref{sec: discuss}. We summarize our conclusions and outline future prospects in Section \ref{sec:Conclusion}. 

We assume a flat, $\Lambda$CDM cosmology
with $\Omega_\Lambda$ = 0.7, $\Omega_M$ = 0.3, and $H_0$ = 67.4 $\pm$ 0.5 km s$^{-1}$ Mpc$^{-1}$ for this work \citep{Planck2018}.

\section{Light-Curve Generation} \label{sec:datgen}
In this section, we review our light-curve generation methods. We first discuss the underlying quasar and binary variability models and then the quasar and binary signal parameter grid, methods for adding host-galaxy light, and application of survey design options in the following subsections.  In this work, we assume that the binary pair of SMBH are embedded within a circumbinary accretion disk \citep[e.g.,][]{BoganovicReview2022, Charisi_2022, DOrazioCharisiBook2023} such that their observations contain stochastic variability from thermal fluctuations in the disk in the same fashion as single quasars \citep{Kelly2009, Kozlowski2010, MacLeod_2010, Zu_2013, Suberlak2021}. A more detailed description of this light-curve generation method with a random, uniform parameter grid and sinusoidal curve fit analysis results can also be found in \cite{Davis_2024}.


To make the light curves we first construct an underlying quasar variability model by assuming a damped random walk (DRW) based on quasar properties \citep{MacLeod_2010, MacLeod2012}. We note that though the DRW is less appropriate on short (hour to day) and poorly constrained on longer (decade plus) timescales \citep[e.g.,][]{Mushotzky_2011, Kozlowski_2016}, it is still a successful empirical model for quasar variability timescales of a month to several years \citep{Suberlak2021}. As this work is a test for general trends when recovering a binary SMBH signal from a model of quasar variability  
that is well-documented to correlate with quasar properties \citep{Suberlak2021}, we continue to use the DRW with the expectation that Rubin/LSST will improve quantitative descriptions and long-term characterization of quasar variability with its long-duration monitoring.
We use the quasar parameters in Table \ref{tab:paramgrid} to estimate the two DRW parameters: the characteristic damping timescale ($\tau$; days) and DRW amplitude (\SFinf; magnitude). The two parameters are estimated using the coefficient prescriptions in \cite{MacLeod2012}. In future work we plan to test other quasar variability models such as the continuous-time autoregressive moving average model \citep{KellyCARMA2014} and the damped harmonic oscillator model \citep{YuRichards2022}. We may also include explicit consideration of the variability contributions from binary mini-disks to explore if this changes our recovery efforts \citep{ArtymowiczLubow1996, MacFadyen_2008, Westernacher_Schneider2022}.

Second, we simulate a simple binary SMBH periodicity model. Physically, we expect periodic binary signals due to orbital motion of the binaries (Doppler boosting) and/or hot spots in the accretion disks. It is also possible that periodic (or ``bursty") accretion and gravitational self-lensing can result in periodic modulation of the light curve. These models are discussed in detail in \cite{Davis_2024} and in the work of \cite{De_Rosa_2019},
\cite{DOrazioCharisiBook2023}, and references therein. We model the binary SMBH periodic signal using a combination of sine and sawtooth functions. To combine the models, we set a parameter called ``sawtoothiness", a percentage of how ``sawtooth-like" the binary signal is. If ``sawtoothiness" is 1.0, the binary signal is a sawtooth signal. If it is 0.0, it is a pure sine. The resulting, convolved model takes the following form:
    \begin{equation}
    \begin{aligned}
        \Delta m(t) = (1-S) * A \sin\left( f t + \phi\right) \\
        +~ S * A\ \textrm{sawtooth} \left( f t + \phi\right) ,
    \end{aligned}
    \end{equation}
where $\Delta m$ is the change in magnitude, $m$, as a function of time $t$ with parameters $A$ (amplitude), $f$ (frequency, $f=2\pi/P$), $\phi$ (phase), and $S$ for the sawtooth parameter percentage \citep{Davis_2024}.
The resulting periodicity model is added to the quasar variability model to make the base light curve.

Third, to investigate host-galaxy contamination of the measured quasar light curves, we contaminate the base light curve with a model for non-variable host-galaxy light. We apply a simple host-galaxy contamination model built from the correlations between quasar mass and host-galaxy luminosity observed by \cite{Li_2023}. This model and the contamination process is described further in Section \ref{sect:Contam} and is featured in Figures \ref{fig:HostGalSumm} and \ref{fig:HostGal}.

Finally, we apply the cadence, seasonal duration, noise properties, and additional observational considerations (like poor observing conditions) to create a fully simulated Rubin light curve. The cadence and light curve length (duration) are further described in Section \ref{sec:SurveyDesign}. Each light curve is resampled within its expected LSST photometric error from \cite{Ivezi__2019}. We add additional flux uncertainty to simulate lunation and random observation conditions (see Section 2.3 of \cite{Davis_2024} for more information). We apply a 7 month seasonal observation duration to every light curve to simulate a target in an equatorial field like the COSMOS and XMM-LSS Deep Drilling Fields. An additional 30\% of the data is randomly removed to simulate data loss from poor observing conditions or engineering time.

\begin{figure*}[t]
\centering
\noindent\includegraphics[width=\textwidth]{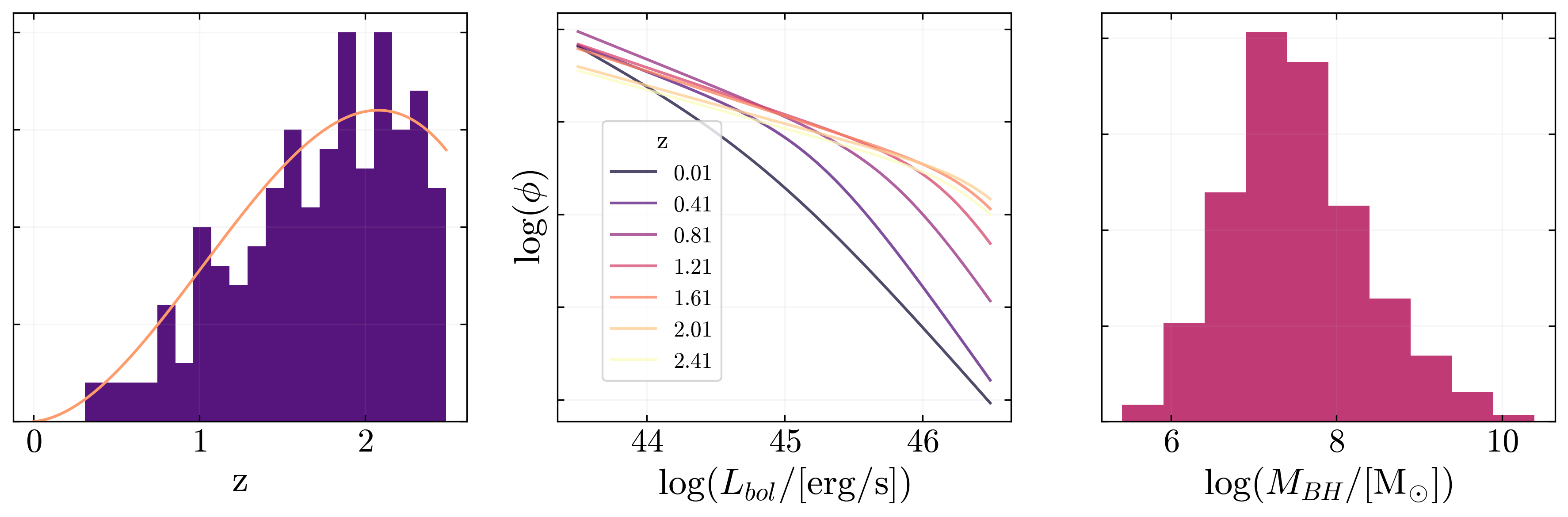}
\caption{A summary of the underlying quasar parameter grid distributions. \textit{Left:} The distribution of randomly drawn redshifts (purple histogram) from the QLF (orange curve). \textit{Center:} The redshift-dependent QLFs that are then sampled, along with a randomly drawn Eddington ratio, to create the distributions of black hole masses, seen \textit{Right}.}\label{fig:ParamgridSum}
\end{figure*}

\begin{figure*}[t]
\centering
\noindent\includegraphics[width=\textwidth]{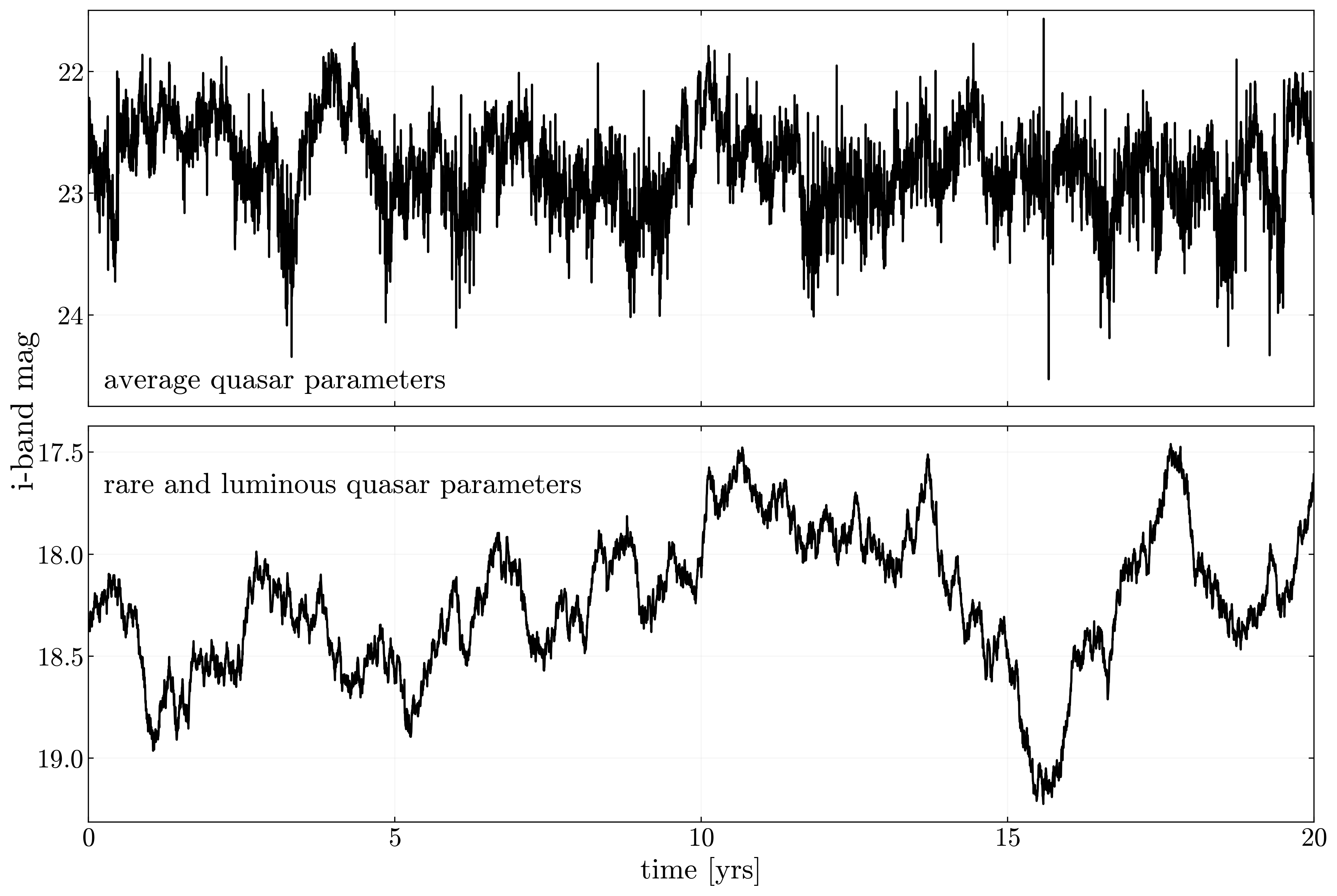}
\caption{Example light curves from the observation-driven quasar parameter grid shown in Figure \ref{fig:ParamgridSum} and derived from Table \ref{tab:paramgrid}. \textit{Top:} A full light curve example simulated from the most common quasar parameters, representing a single quasar at z $\sim$ 1.72 with a mass of $10^{7.52}$ $M_{\odot}$ and a bolometric luminosity of $10^{45.7}$ ergs/s. \textit{Bottom:} A full light curve example of a rare and luminous single quasar at z $\sim$ 0.76 with a mass of $10^{9.79}$ $M_{\odot}$ and a bolometric luminosity of $10^{46.2}$ ergs/s.}\label{fig:ExampleLCs}
\end{figure*}

\begin{figure*}[t]
\centering
\noindent\includegraphics[width=\textwidth]{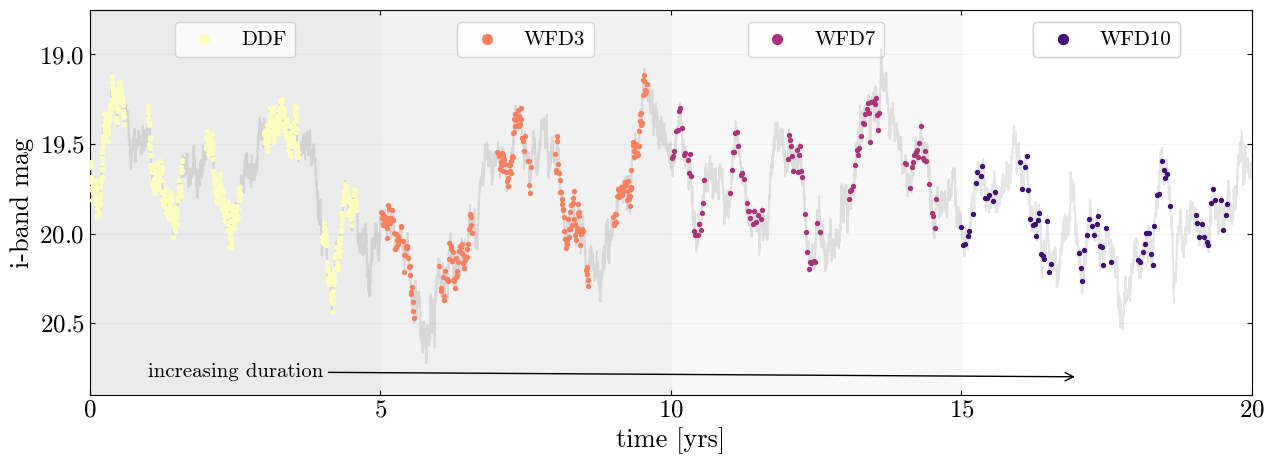}
\caption[Rubin Survey Design Cadence and Duration Combinations]{Survey design choices of cadence and duration over an example light curve (solid light gray line). The model light curve is of a binary SMBH system at z $\sim$ 1.4 with a total mass of $10^{9.49}$ $M_{\odot}$, bolometric luminosity of $10^{46.1}$ ergs/s, binary amplitude of 0.16 mag, and period of 5325 days. The gray background gradient delineates 5 yr increments for survey duration. 
The cadences choices are depicted as the scattered points on top of the model light curve, from most to least dense observation schedule. A 1-day survey cadence for the DDFs is shown in the first 5 yrs (yellow scatter; darkest gray box). The next survey cadences are for the WFD main survey, representing a 3-day (WFD3; orange scatter), 7-day (WFD7; pink scatter), and 10-day (WFD10; purple scatter) cadence. Seasonal observation gaps and random weather loss are also included in each cadence depiction. There are 16 total combinations of cadence and duration explored in this work.}\label{fig:SurveyDesign}
\end{figure*}

\subsection{ Binary and Quasar Signal Parameters}\label{sec:paramgrid}
The parameter grid was designed to sample the expected population of quasars observed by Rubin/LSST. The parameter grid, seen in Table \ref{tab:paramgrid}, was used to generate two, separate datasets used in this work: a non-contaminated (quasar-only) and a host-galaxy contaminated dataset. After an applied $i$-band magnitude cut of 23.5, the non-contaminated dataset had 1.96 million light curves and the host-galaxy contaminated dataset had 2.93 million. As described in \cite{Davis_2024}, the magnitude cut is applied to preserve light curves with sufficient signal-to-noise quality when utilizing the predicted noise model for Rubin photometry \citep{Ivezi__2019}.

To build the suite of model light curves, we start with the expected LSST AGN number density at a given redshift from Table 10.2 of \cite{LSST2009}, derived from the quasar luminosity function found in \cite{Hopkins2007}. The expected distribution is shown as the solid orange line in the left-most panel of Figure \ref{fig:ParamgridSum}.  We limit our redshift distribution to $0 < z  < 2.5$ to cover the peak and majority ($\sim60\%$) of the AGN number density expectation and to avoid Ly$\alpha$ forest absorption in the bluest LSST filter.
We then sample the expected AGN redshifts 200 times, shown as the purple histogram in the left-most panel of Figure \ref{fig:ParamgridSum}. 

The quasar parameters in our parameter grid are then derived from the UV/optical, redshift-dependent, double power-law bolometric quasar luminosity function (QLF), Equation 13, from \cite{Shen2020_QLFs},
\begin{equation}
\frac{\mathrm{d} n}{\mathrm{~d} M}=\frac{\phi_{*}^{\prime \prime}}{10^{0.4(\alpha+1)\left(M-M_{*}\right)}+10^{0.4(\beta+1)\left(M-M_{*}\right)}},
\end{equation}
where $\phi_{*}^{\prime \prime}$ is
the co-moving number density normalization, $M_{*}$ is the break magnitude, and $\alpha$ and $\beta$ are the faint-end and bright-end slopes. The redshift-dependent values of each parameter were set to the ``local `polished'" QLF fit results of Table 3 in \cite{Shen2020_QLFs}. Example QLF curves for fixed redshifts are shown in the center panel of Figure \ref{fig:ParamgridSum}.  The QLF curve at each redshift is used as a probability distribution to draw 200 sampled bolometric luminosities. Each luminosity then gets 200 samples from a simple log-linear probability density (slope of 1) of Eddington ratios that reflects the assumption that small $L/L_{Edd}$ dominate over large $L/L_{Edd}$ \citep{KellyShen_2013}. The black hole masses are computed from the bolometric luminosity and Eddington ratio. The distribution of quasar masses can be seen in the right-most panel of Figure \ref{fig:ParamgridSum}. The top and bottom panels of Figure \ref{fig:ExampleLCs} are example light curves that represent the average quasar parameter pairing (top) and an uncommon parameter pairing (bottom). 

The binary parameters for each model light curve are random choices from uniform distributions between their bounds listed in Table \ref{tab:paramgrid}.  The binary amplitude is limited from $0 < A < 0.75$ mag where a binary amplitude of zero indicates a single quasar.  The binary period ranges from 0.5 yrs to 15 yrs. The ranges were chosen such that they represent close-separation binaries that are unlikely to significantly evolve over the Rubin survey duration due to their long lifetimes and had at least 1.5x period cycles represented in the light curve duration at 20 yrs. Allowing for longer ($>10$ yr) binary periods allows us to investigate recovery potential and confusion sources of periods longer than the expected LSST baseline. The long binary periods are still injected into short ($5$ yr) baselines to test early binary recovery capabilities.

\begin{table}[!ht]
\centering
\begin{tabular}{ l c  }
\hline 
Parameter & \makecell{Distribution,  \\ bounds [lower, upper]} \\
\hline
log($M/M_{\odot}$) & QLF, [5.4, 10.4] \\ 
log($L_{bol}/$[ergs$/$s]) & QLF, [43.5, 46.5]  \\ 
z & QLF, [0, 2.5] \\ 
amplitude [mag] & Random uniform, [0.0, 0.75] \\ 
period [days] & Random uniform, [180, 5400]  \\ 
sawtoothiness [\%] & Random uniform, [0.0, 1.0] \\ 
phase &  Random uniform, [0.0, 2$\pi$]  \\
\hline
\end{tabular}
\caption[An Updated Light Curve Parameter Grid]{The range of quasar properties of (total) black hole mass, luminosity, and redshift and binary signal parameters of amplitude, period, signal type, phase used to generate the simulated light curves. A light curve with a zero amplitude binary signal is considered a single quasar light curve. The quasar luminosities are calculated from the UV/optical bolometric quasar luminosity function (QLF) of \cite{Kulkarni2019}. The masses are then derived from these luminosities with a linear assumption that lower Eddington ratios dominate over higher Eddington ratios. The redshifts are representative of the expected AGN redshifts to be observed by Rubin as depicted in Table 10.2 of \cite{LSST2009}, which are derived from the QLF of \cite{Hopkins2007}, that is the basis for \cite{Kulkarni2019}. We randomly sample each binary parameter from a uniform distribution within its stated range.}
\label{tab:paramgrid}
\end{table}

\subsection{Cadence and Baseline}\label{sec:SurveyDesign}
We test binary SMBH detection over the 10 yr lifetime of LSST and beyond to 20 yrs. We also investigate cadence requirements for binary SMBH recovery with sinusoidal curve fits in the different Rubin survey regimes. Each dataset of light curves is modeled, fully, up to 20 yrs and then cut to the desired baseline during the analysis stage. We model four light curve baselines of 5, 10, 15, and 20 yrs. We apply four cadence possibilities that include effective cadences of 1-day (Deep Drilling Field; DDF) and three main survey cadences of 3-day (WFD3), 7-day (WFD7), and 10-day (WFD10). There are 16 total combinations of each cadence and duration option. These options are depicted in the survey design summary figure, Figure \ref{fig:SurveyDesign}.  Future work includes the full implementation of the \cite{Bianco_2022LSST} rolling survey cadences of staggered visits in each filter. 

\subsection{Host-Galaxy Contamination}\label{sect:Contam}
\begin{figure*}[t]
\centering
\noindent\includegraphics[width=\textwidth]{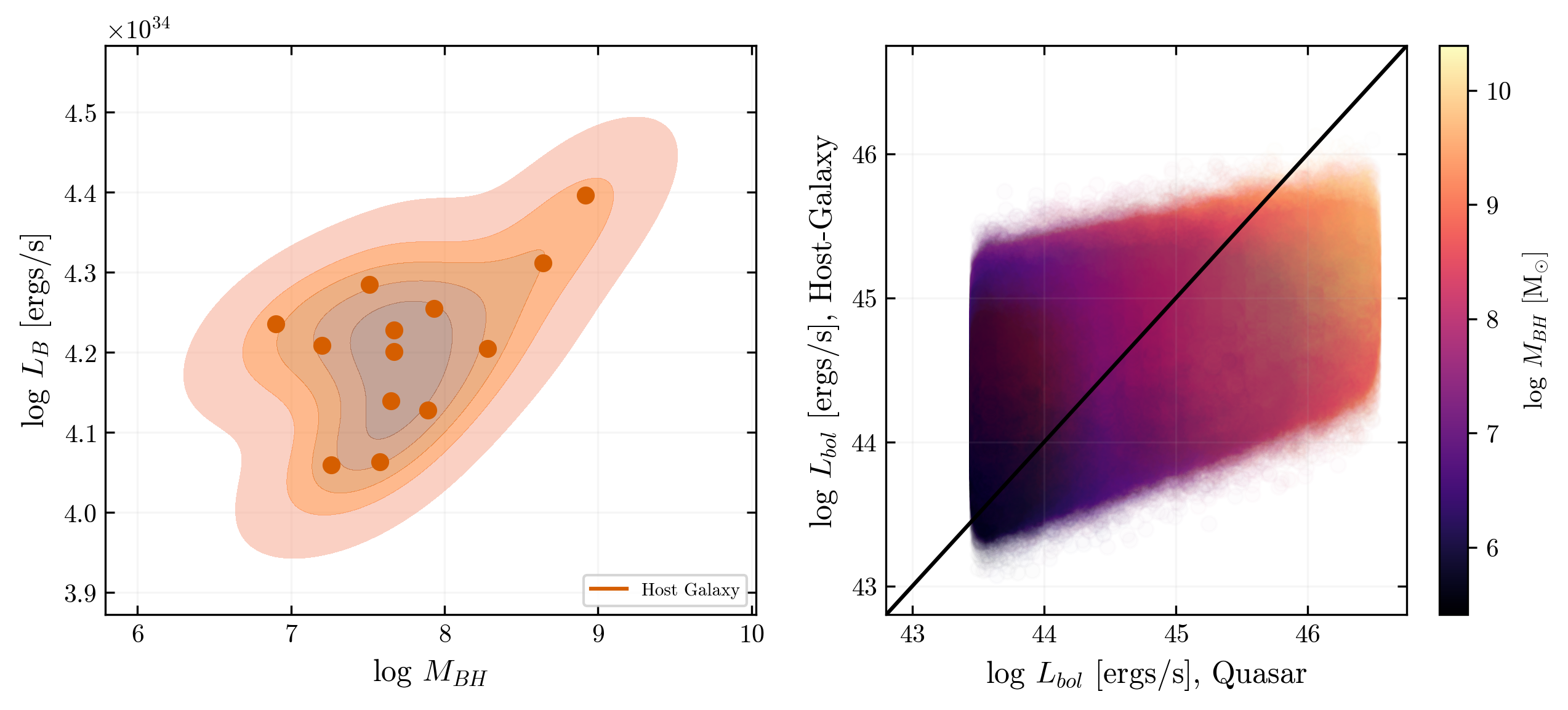}\caption{A summary on the host-galaxy light contamination model construction. \textit{Left:} The probability density distribution (orange contours) of host-galaxy $B$-band luminosities of spectroscopically confirmed SDSS-RM quasars in terms of their quasar mass, from Table 3 of \cite{Li_2023} (orange points). This distribution was sampled to create the host-galaxy contamination model for our single and binary quasar light curves. \textit{Right:} The resulting parameter grid distribution of host-galaxy and associated quasar bolometric luminosities, color-coded by quasar mass. A 1:1 ratio line is plotted to highlight that most light curves in our distribution will be quasar-dominated (below the line) and not host-galaxy dominated (above the line). Quasars with lower black hole masses are more likely to have larger host-galaxy luminosities in comparison to their quasar luminosity.}\label{fig:HostGalSumm}
\end{figure*}

\begin{figure*}[t]
\centering
\noindent\includegraphics[width=\textwidth]{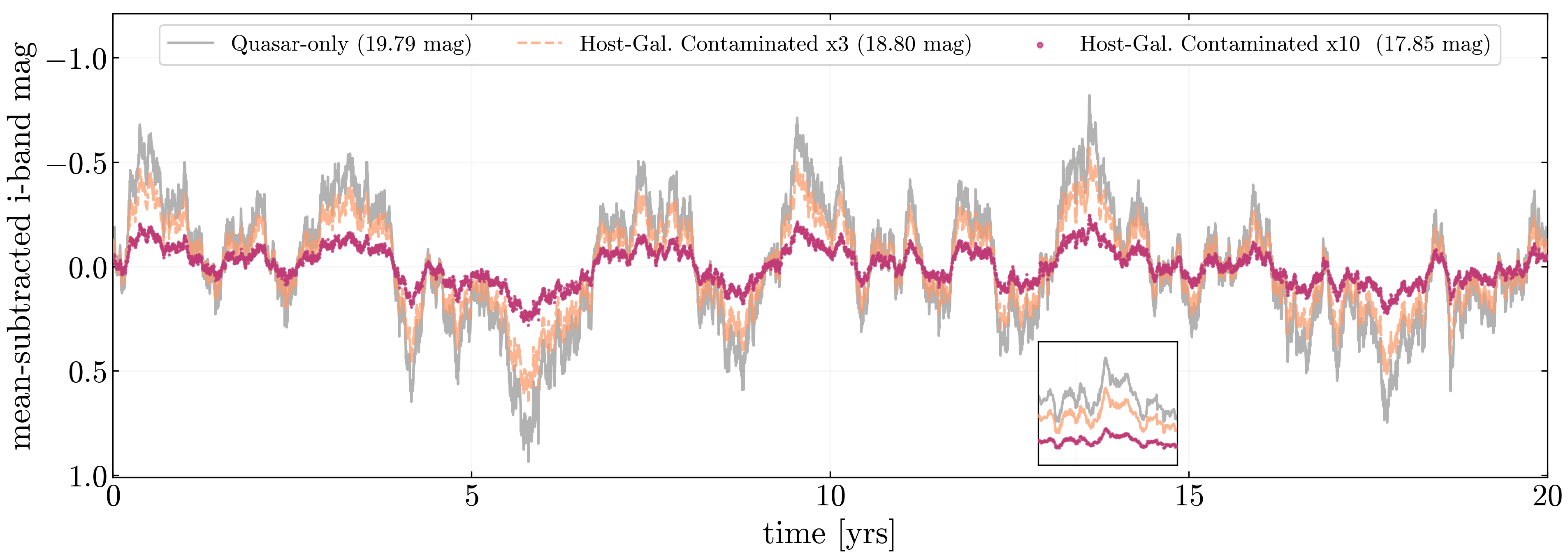}
\caption[Examples of the Implemented Host-Galaxy Contamination Procedure]{Examples of the implemented host-galaxy contamination procedure, using a model light curve of a binary SMBH system at z $\sim$ 1.4 with a total mass of $10^{9.49}$ $M_{\odot}$, bolometric luminosity of $10^{46.1}$ ergs/s, binary amplitude of 0.16 mag, and period of 5325 days. Mean-subtracted examples of 3$\times$ (orange dashes) and 10-times (pink scatter) host-galaxy contamination compared to the quasar-only light curve (gray solid line). The host galaxy light results in weaker apparent variability due to dilution from the constant-brightness galaxy emission.}\label{fig:HostGal}
\end{figure*}

We estimate the effects of host-galaxy contamination on binary classification. To contaminate the light curves, we first create a model for host galaxy brightness as a function of black hole mass. 
Shown in the left panel of Figure \ref{fig:HostGalSumm} is the distribution of the host-galaxy, $B$-band luminosities and quasar masses of spectroscopically confirmed SDSS-RM quasars of Table 3 from \cite{Li_2023}, which explored BH-galaxy correlations in a similar quasar parameter space (redshift and luminosity) to our simulations using BH masses derived from reverberation mapping \citep[RM;][]{1993RM, Cackett_2021} and host-galaxy light contributions calculated from HST imaging and 2D image decomposition via GALFIT \citep{2010GALFIT}. We sample this distribution to calculate estimates of the host-galaxy bolometric luminosity based on the quasar mass of each parameter grid pairing. The right panel of Figure \ref{fig:HostGalSumm} depicts the bolometric luminosities of our quasars and their estimated host-galaxy components. 
We introduce host-galaxy contamination by adding constant host-galaxy light to each light curve. We then reduce the light curve standard deviation by utilizing a multiplicative factor calculated by the ratio of host-galaxy to total (quasar+host) light. For example, if the host-galaxy luminosity is twice that of the quasar, the standard deviation is reduced by a factor of $2/3$.
In our simplified paradigm, the host-galaxy contamination effectively dilutes the apparent light curve variability since the variable quasar light represents less of the total flux.

Figure \ref{fig:HostGal} summarizes this process. 
As LSST progresses, we will be able to develop effective methods of quantifying and removing host-galaxy contamination from quasar light curves to minimize general contamination, most notably via combined LSST-Euclid SED fitting.

\section{Light Curve Fitting}\label{sect:Analysis}

We fit the light curves with a simple sinusoidal function to describe the periodic variability. This fitting technique is fast and computationally efficient and so can be effectively used as a rapid method of light curve assessment (e.g. for event brokers). The details of the fitting method are described in \cite{Davis_2024}.
In brief summary, we utilized the Python package \texttt{LMFIT}, a non-linear least-squares minimization and curve-fitting tool for Python that utilizes the Levenberg–Marquardt (robust, damped least-squares) algorithm, and its built-in \texttt{SineModel} for the curve fits \citep{lmfits}. We determine goodness-of-fit using the reduced chi-squared statistic, \redchi. Of note, we assign an error floor of 0.5\% the mean of the fitted light curve to avoid over-fitting bright quasars that are expected to have low photometric error with Rubin. See \cite{Ivezi__2019} for more information on the expected Rubin photometric error and \cite{Davis_2024} for the detailed implementation, testing, and results of the initial application of the curve fits used here. 

The goal of curve fitting the data was to test recovery of the binary parameters, amplitude, phase, and period, from the simulated DRW plus binary light curves. We fit the same sine function to the pure-DRW, single-quasar light curves to estimate the binary false-positive detection rate due to quasar variability.
The curve fits are also a useful, computationally inexpensive method that can provide fast insight into binary parameter recovery and false-positive rates. Through the implementation of basic parallelization, we are able to fit $\sim$42,000 light curves per minute compared to $\sim$7,500 fits per minute in \cite{Davis_2024}. This method can be used as an initial triage step before applying more computationally intensive methods \citep[e.g.,][]{Witt_2022, Robnik2024}.

\section{Results}\label{sec: results}
Here we present the results of potential binary SMBH discovery with Rubin as a function of quasar properties, numerous survey baseline and cadence considerations, and inclusion of host-galaxy light. 

In general, the false-positive detection is defined as quasar-only light curves with fit results of a \redchi $< 5$ and reasonable binary amplitude ($A$) and period ($P$) fit values. We define reasonable binary fit amplitudes as $A > 0.1$ so that the binary amplitude can be distinguished from the quasar variability amplitude and reasonable binary fit periods as $180 < P < $ duration/1.5 so that the monitoring duration includes at least 1.5 period cycles. 

We generally define the binary parameter recovery as binary light curves with a \redchi $< 5$ and the fit result values within 10\% of the input binary parameter. For simplicity, we use $\Delta X$/$X_{in}$, where $\Delta X = X_{fit\ results} - X_{in}$ for parameter recovery calculations. These definitions were established in and are discussed more in \cite{Davis_2024}.

\subsection{Binary Detection as a Function of Quasar Properties}\label{sec:ParamGridResults}
\begin{figure*}
\centering
\noindent\includegraphics[width=\textwidth]{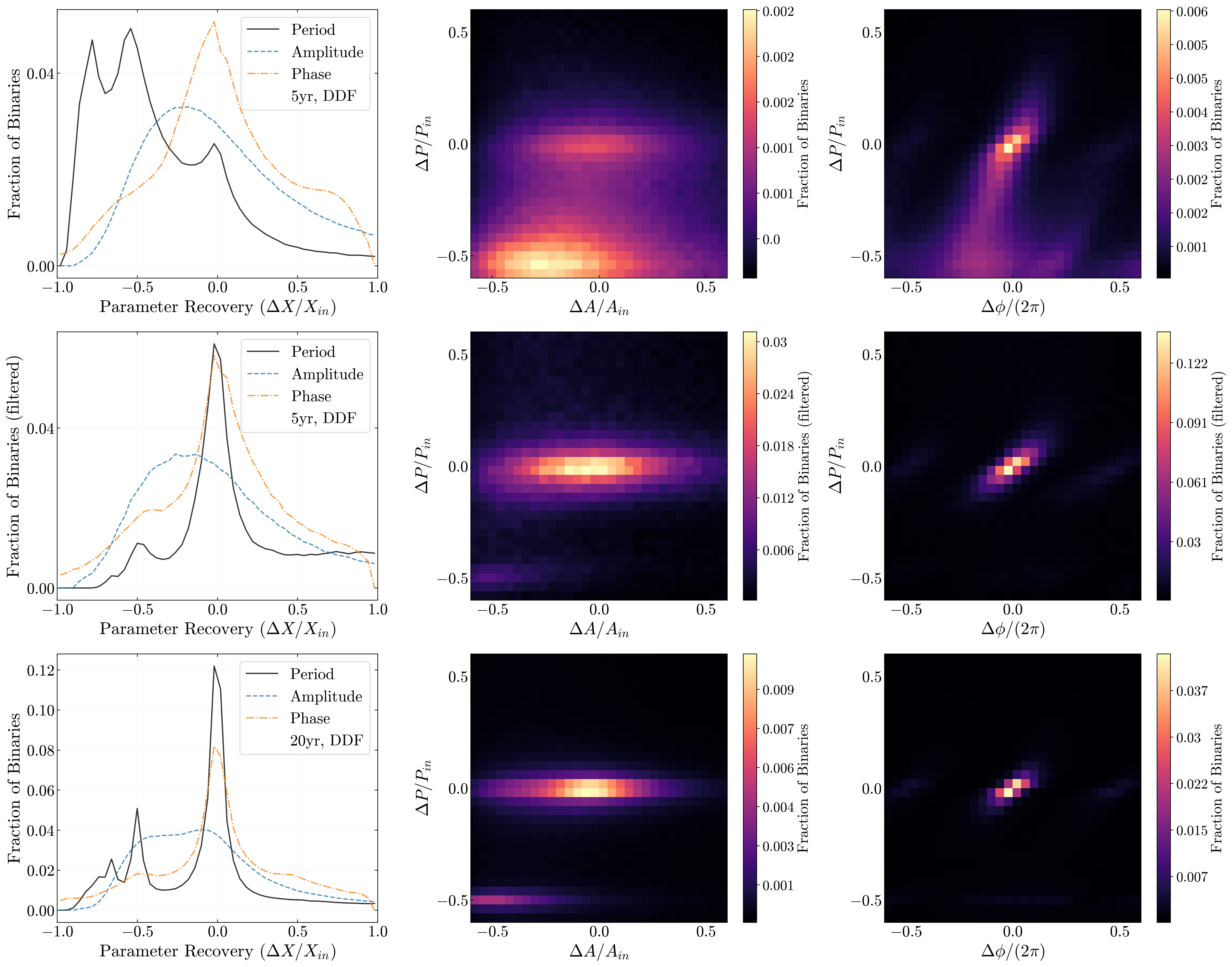}
\caption[The Fractional Binary Parameter Recovery]{The fractional binary parameter recovery for amplitude ($A$), period ($P$), and phase ($\phi$), using curve fits. 
\textit{Top:} the 5-yr DDF (1-day cadence) light curve all-parameter recovery results (left), period versus amplitude recovery (middle) and period versus phase (right). Period recovery for these light curves is poor and typically underestimated by a factor 2.  \textit{Middle:} The 5-yr, DDF light curves but excluding periods longer than 3.3 (duration/1.5) yrs. The binary period is much more likely to be reliably recovered when it is less than the monitoring duration. \textit{Bottom:} Binary recovery for the 20-yr DDF light curves, unfiltered. Most periods are effectively recovered but with a significant population of 2$\times$ underestimated periods associated with particular light curve shapes, as shown in Figure \ref{fig:SawtoothinessSpikeInvest}. }\label{fig:EnsembleRecov}
\end{figure*}

\begin{figure}
\centering
\noindent\includegraphics[width=\linewidth]{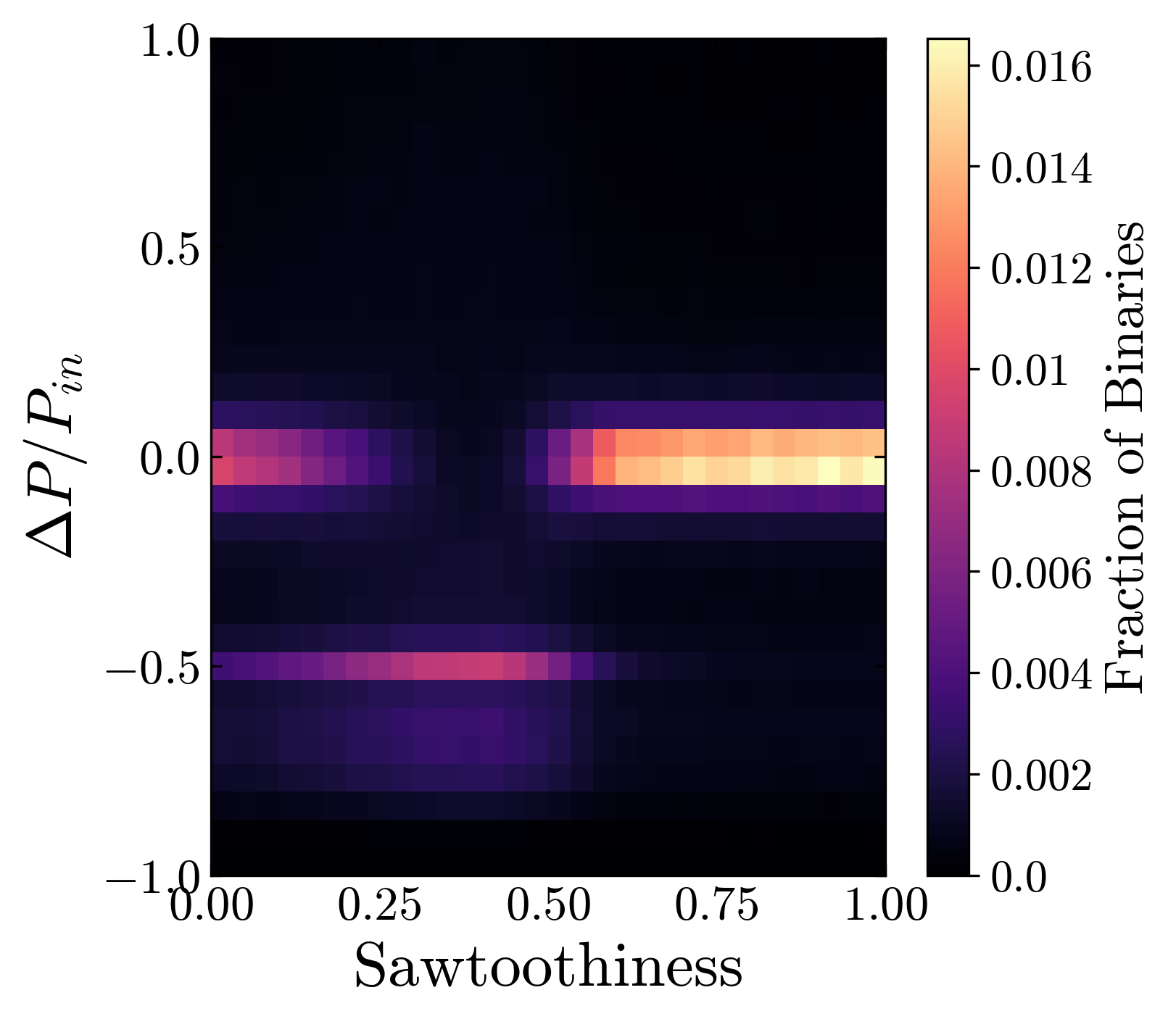}
\caption[Period Recovery as a Function of ``Sawtoothiness"]{Period recovery as a function of ``sawtoothiness".
Sawtooth-like binary signals (sawtoothiness $>0.5$) produce extreme flux deviations, which can be disentangled from the quasar variability and result in better binary recovery. In contrast, smoother sinusoidal variability (sawtoothiness $<0.5$) is more likely to be confused with quasar DRW variability, which frequently results in best-fit periods that are 2$\times$ lower than the input. The poor period recovery for binaries with moderately low sawtoothiness is responsible for the $\Delta P/P_{in}  \approx 0.5$ spikes seen in Figure \ref{fig:EnsembleRecov}.
}\label{fig:SawtoothinessSpikeInvest}
\end{figure}

\begin{figure*}
\centering
\noindent\includegraphics[width=\textwidth]{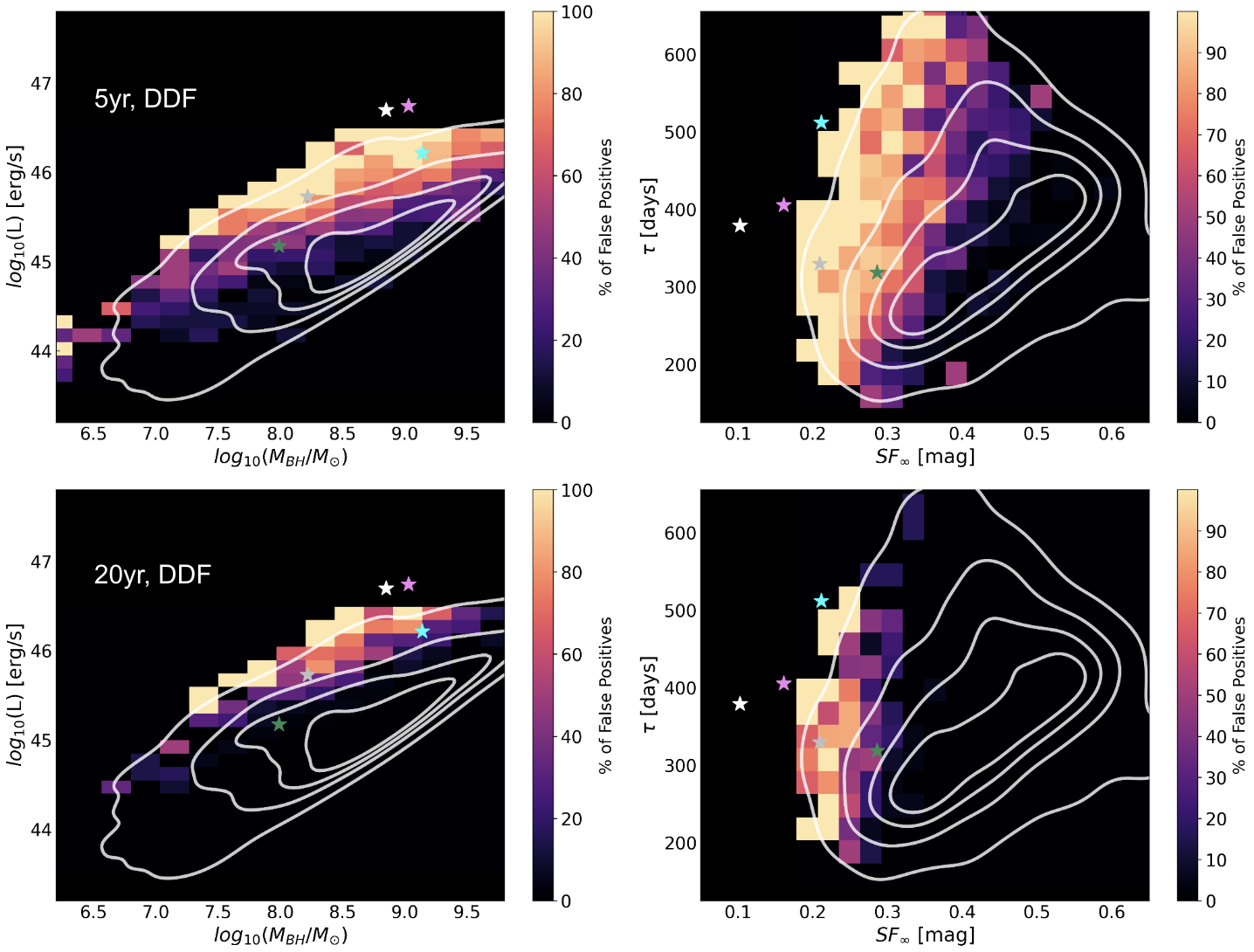}
\caption{2-D histograms of the false-positive demographics, contours (white) of the binary population, and stars of five, real photometric binary candidates from \cite{Graham2015}, collected in Table 3 of \cite{Davis_2024}. Results are shown for the 5-yr DDF (top row) and 20-yr DDF (bottom row) light curves. \textit{Left panels:} False-positive rates in terms of quasar luminosity and mass. Regardless of light curve duration, the highest luminosity light curves in our datasets feature high false-positive rates. As duration increases, the false-positive rate decreases. \textit{Right panels:} False-positive rates in terms of DRW parameters, damping timescale $\tau$, and quasar variability amplitude \SFinf. We find that low values of quasar variability amplitudes at most values of  damping timescale dominate the false-positive demographics. }\label{fig:FPratesVSQuasarProperties}
\end{figure*}

\begin{figure*}
\centering
\noindent\includegraphics[width=\textwidth]{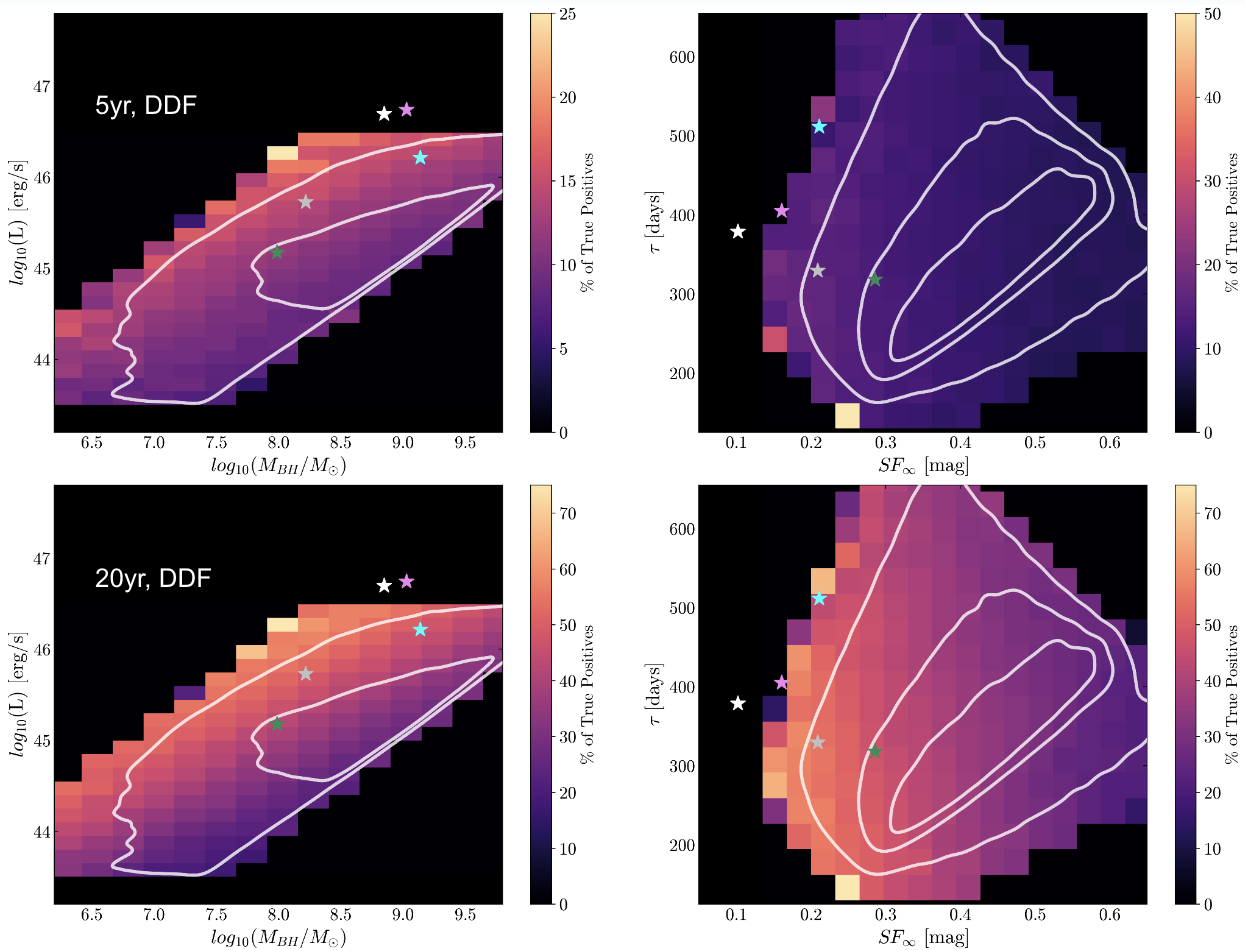}
\caption{2-D histograms of the true-positive, binary period recovery demographics, contours (white) of the binary population, and stars of five, real photometric binary candidates from \cite{Graham2015}, collected in Table 3 of \cite{Davis_2024}. Results are shown for the 5-yr DDF (top row) and 20-yr DDF (bottom row) light curves. \textit{Left panels:} True-positive rates in terms of quasar luminosity and mass. \textit{Right panels:} True-positive rates in terms of DRW parameters, damping timescale $\tau$, and quasar variability amplitude \SFinf. The most luminous systems produce the highest true-positive binary period recovery rates, with a large increase in recovery as light curve baseline increases.}\label{fig:TPratesVSQuasarProperties}
\end{figure*}

Figure \ref{fig:EnsembleRecov} depicts the binary parameter recovery with a realistic quasar property parameter grid and a large range in binary periods. The top row shows that for 5 yr light curves with a 1-day effective DDF cadence, binary parameter recovery is poor and is notably underestimated for the binary period. If we filter out the input binary periods that have less than 1.5 cycles in the light curve duration, as shown in the middle row, all parameter recovery improves. Curve fits are not effective at recovering periods longer than the light curve duration. 
The bottom row of panels in Figure \ref{fig:EnsembleRecov} shows the binary parameter recovery if the light curve duration is extended to 20 yrs without any period filter. 
The amplitude is poorly constrained and generally underestimated. The period and phase binary parameters are typically recovered in tandem with each other. Notably, there is a significant peak in recovered periods with $\Delta P/P_{in} \approx -0.5$ in the 20 yr data, which is also evident in the 5 yr data.

Figure \ref{fig:SawtoothinessSpikeInvest} investigates period recovery as a function of signal type (``sawtoothiness"). Values of sawtoothiness roughly between 0.2 and 0.5, where the signal is a mixture of pure sine and sawtooth, are poorly recovered and constitute the spikes in period recovery at $\Delta P/P_{in} \approx -0.5$. These periods are underestimated and are reported to be near a quarter of the light curve duration. Per Anderson-Darling \citep[AD;][]{AndersonDarling1952} tests, where we compared the distribution of parameters that constitute the period-recovery spikes to their parent parameter grid distributions, sawtoothiness is the only parameter to cause the period recovery deviations. The rest of the possible parameters do not deviate from their parent distributions.

We investigate the false-positive rate in terms of quasar mass and luminosity in Figure \ref{fig:FPratesVSQuasarProperties} for 5-yr and 20-yr DDF light curves. Regardless of light curve duration, the false-positive rate is dominated by high-luminosity quasars. We show the same results for true-positive binary recovery detections, in Figure \ref{fig:TPratesVSQuasarProperties}.  We also find that as duration increases, the false-positive rate is suppressed for lower luminosity quasars. At the same time, the true-positive rate improves for high-luminosity quasars. For 5-yr DDF light curves, the global false-positive rate is 14.3\%. For the 20-yr DDF light curves, the false-positive rate declines to 5.8\%.


\begin{table*}[!ht]
\centering
\begin{tabular}{ l |c| c c c c }
\hline 
\textbf{Duration} (yr) & False-Positive & & Recovery (\%)&  \\
and Cadence & Rate (\%) & $P$ & $A$ & $\phi$ & All  \\
\hline 

\textbf{5} &&&&&\\
WFD10	&14.04	&10.59	&14.45	&22.68 & 1.26\\
WFD7	&14.80	&10.60	&14.45	&22.59 & 1.25\\
WFD3	&14.69	&10.77	&14.52	&22.70 & 1.30\\
DDF 	&14.31  &10.83  &14.56  &22.71 & 1.32\\
\textbf{10} &&&&&\\
WFD10 &	8.86	&23.29	&16.63	&29.22 & 3.80\\
WFD7	&9.42	&23.51	&16.66	&29.27 & 3.84\\
WFD3	&9.80	&23.89	&16.81	&29.44 & 3.93\\
DDF	&9.95	&24.10&	16.82&	29.54 & 3.98\\
\textbf{15} &&&&&\\
WFD10   &6.62	&29.33	&17.39	&30.47 & 5.56\\
WFD7	&7.34&	29.63	&17.47	&30.58 & 5.63\\
WFD3	&7.27	&30.16	&17.64	&30.78 & 5.79\\
DDF	    &7.34	&30.48	&17.69	&30.90 & 5.85\\
\textbf{20} &&&&&\\
WFD10 &	5.45	&34.97	&17.84	&31.56 & 7.07\\
WFD7	&5.64&	35.39	&17.96	&31.66 & 7.18\\
WFD3  &5.64	&36.06	&18.15	&31.95 & 7.37\\
DDF	& 5.83	& 36.46 &18.26	&32.10 & 7.48\\
\hline
\end{tabular}
\caption{The false-positive and binary parameter recovery rates for all 16 combinations of cadence and duration explored in this work. Generally, these rates are affected more by a change in duration than a change in cadence. As duration increase, false-positive rate decreases and recovery rates increase. The most notable rate of change occurs as light curve durations increases from 5 yrs to 10 yrs. }
\label{tab:resultsMega}
\end{table*}

\subsection{Binary Detection as a Function of Survey Design}\label{sec:CadenceAndDurationlResults}
The general false-positive rate and binary parameter recovery values for all combinations of light curve duration and cadence are presented in Table \ref{tab:resultsMega}. The binary amplitude is the worst-recovered parameter. The recovery of binary phase is relatively strong and stable compared to the two parameters. The binary period is the parameter most sensitive to survey design, discussed further below.  

\begin{figure*}
\centering
\noindent\includegraphics[width=\textwidth]{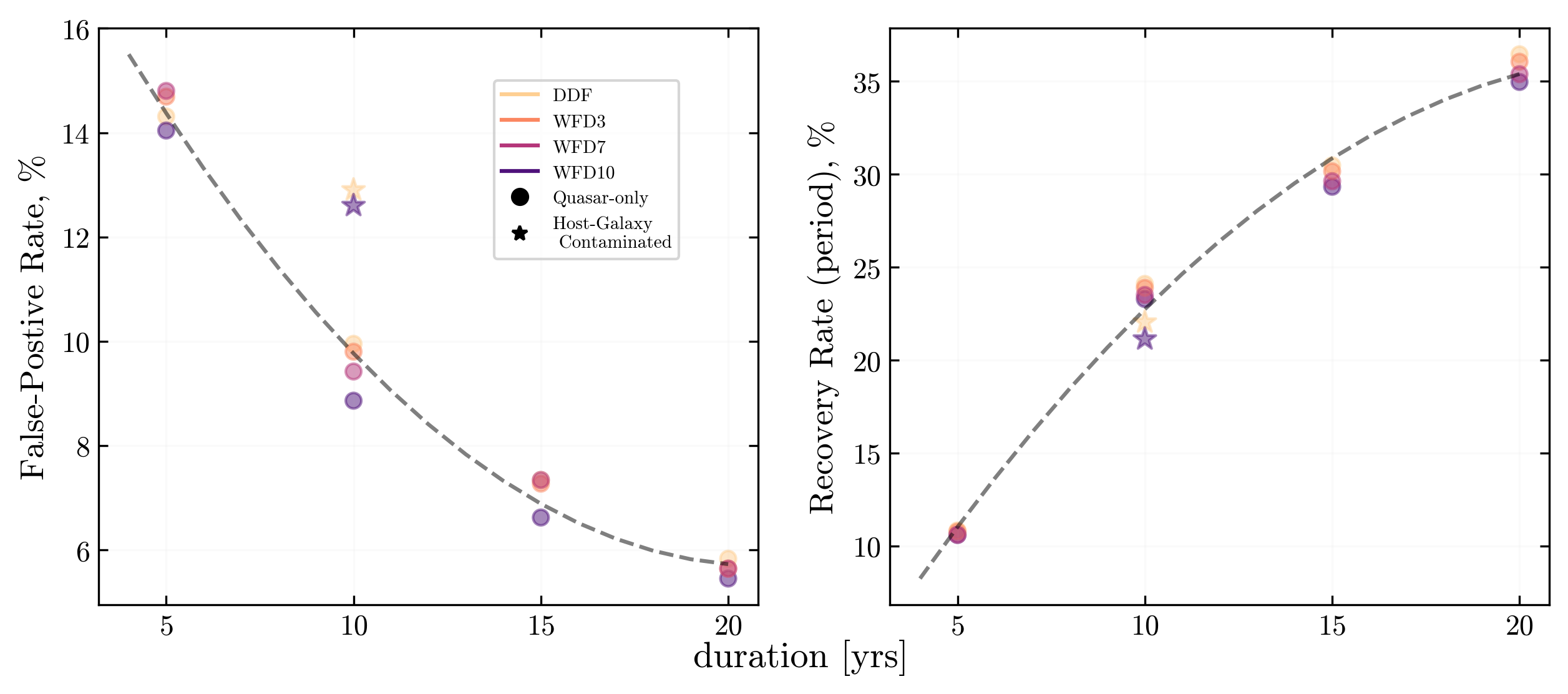}
\caption[The False-Positive and Period Parameter Recovery Rates as Functions of Survey Design]{The false-positive rate and period parameter recovery rate as functions of survey duration, color-coded by cadence. \textit{Left:} The false-positive rate from single quasar light curves as a function of light curve baseline (duration), color-coded by cadence. The false-positive rate of uncontaminated, quasar-only light curves (circle marker) decreases as the survey duration increases. The host-galaxy contaminated light curves (star marker) feature higher general false-positive rates because of the light curve smoothing the contamination produces. \textit{Right:} The period parameter recovery rate from binary light curves as a function of light curve baseline (duration), color-coded by cadence.  The recovery rate increases as duration increases. The results for parameter recovery are similar for both host-galaxy contaminated and uncontaminated light curves. For both rates, duration is significantly more important than cadence, observing at the full LSST survey duration of 10 yrs gives the largest rate of improvement over 5 yrs prior, and there is likely no more benefit for short ($< 15$ yr) period binary recovery nor false-positive rate after 20 yrs.}\label{fig:ratesVbaseANDcad}
\end{figure*}

Figure \ref{fig:ratesVbaseANDcad} depicts false-positive total and period recovery rates as a function of light-curve duration, color-coded by cadence.
There are minor ($<$ 5\%) differences when comparing the same duration with different applied cadences for all of these rates. The cadence with the least visits (WFD10) is still largely sufficient for binary detection. It is possible that injecting shorter ($< 0.5$ yr) binary periods could result in differences across cadence with this detection method, which could be explored in future work.

As baseline increases, all of the rates we explore improve (decreasing false-positive rates and increasing recovery 
rates). Binary period recovery is the most-improved binary parameter when duration increases ($<$ 25\% increase), followed by phase ($<$ 8.5\%), and then amplitude ($<$ 5\%). The largest rate of improvement occurs when the light curve duration increases from 5 to 10 yrs. Long light curve durations will be critically important for reliably measuring the binary period.
From Figure \ref{fig:FPratesVSQuasarProperties} we see that the lower-luminosity quasars benefit from false-positive rate suppression in longer light curve baselines.
Longer light curve durations increase the true-positive binary recovery rates for high-luminosity light curves as shown in Figure \ref{fig:TPratesVSQuasarProperties}. Light curve baseline has a larger impact on binary detection rates than cadence.

\subsection{Inclusion of Host-Galaxy Light}\label{sec:HostGalResults}
\begin{figure*}
\centering
\noindent\includegraphics[width=\textwidth]{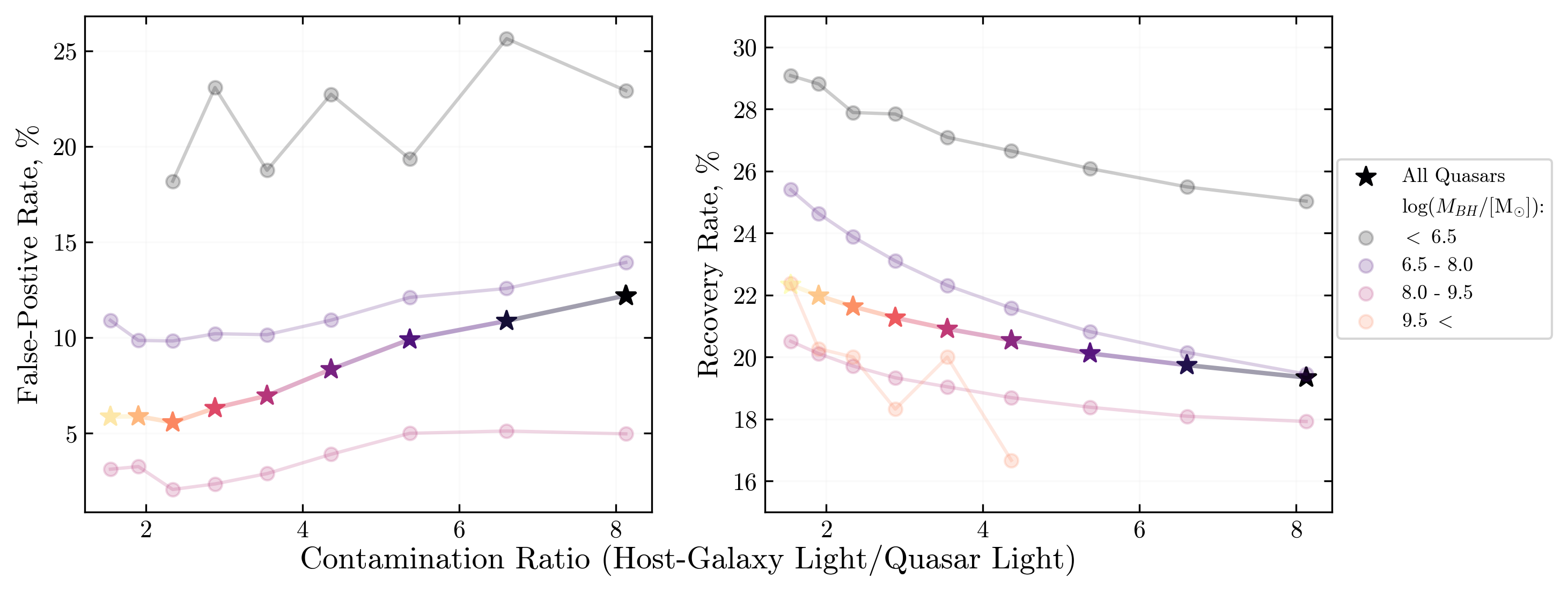}
\caption[The False-Positive and Period Parameter Recovery Rates as Functions of Host-Galaxy Contamination]{The false-positive rate and period parameter recovery rate as functions of host-galaxy contamination, color-coded by quasar mass. \textit{Left:} The false-positive rate of single quasar light curves as a function of contamination. For all quasar masses (star markers; the lighter the marker color, the higher the quasar mass average in that bin), the false-positive rate is positively correlated with contamination ratio. Lower mass quasars typically have higher false-positive rates due to higher host-galaxy contamination. Lower-mass quasars tend to have some degree of host-galaxy contamination in our work. Bins with poor sampling ($<$20 sources) are not depicted. \textit{Right:} The recovery rate of binary light curves as a function of contamination. Generally for all quasar masses (star markers), the recovery rate is negatively correlated with the contamination ratio.}\label{fig:ratesVcontamination}
\end{figure*}

Also plotted in Figure \ref{fig:ratesVbaseANDcad} are reference host-galaxy results, where we see higher false-positive rates due to the contamination processes lowering the standard deviation of contaminated light curves, resulting in better \redchi values.
Figure \ref{fig:ratesVcontamination} shows that as the host-galaxy contamination rate increases, the false-positive rate increases and the period recovery rate decreases for all masses of quasars (star markers). As light curves become more contaminated, their fractional variability amplitude decreases and can produce better \redchi fit result values, resulting in higher false-positive rates with a detection method like simple sinusoidal curve fits. The smoothing effect of host-galaxy contamination can also obscure the binary signal, resulting in lower recovery rates. In our work, the lowest mass quasars, log($M_{BH}/M_{\odot}$) $<$ $6.5$, feature the highest rates of host-galaxy contamination. The low-mass quasars consequently feature higher false-positive and lower recovery rates when host-galaxy contamination is considered.

\section{Discussion} \label{sec: discuss}
In the following section, we discuss the results of applying simple sinusoidal curve fits to synthetic LSST light curves with an observation-driven quasar property parameter grid, host-galaxy contamination, and several survey cadence and duration considerations.

\cite{Davis_2024} established that these curve fits can be generally effective for strong binary signals with amplitudes larger than 0.5 mag and for low-luminosity quasars. In this work, we maintain that they are sufficient for the recovery of binary period and phase in long-duration light curves. We note that the change in injected binary periods (larger range) and amplitudes (smaller in magnitude) compared to \cite{Davis_2024}, when coupled with new quasar properties, produce lower false-positive rates and lower recovery rates. This is due to less representation of high-mass, high-luminosity quasars, which have smaller quasar variability amplitudes and light curve standard deviations. High-mass, high-luminosity quasars produce lower \redchi with simple sinusoidal curve fits and show increased false-positive rates and recovery rates.

As in \cite{Witt_2022}, we find that duration matters more than cadence for general improvement of binary recovery and false-positive rate suppression. Overall, this is a welcome result as binary detection does not require a DDF cadence. We will be able to detect binary SMBHs in the DDFs and in the main survey WFD monitoring field(s) with the LSST 10 yr duration, improving the chances of finding these rare objects. We highly recommend the continuation of observational monitoring of binary SMBH candidates by extending survey observation up to 20 yrs or by exploiting archival observations, when available. The false-positive rate is suppressed at this duration, and it will allow for reliable recovery of binary periods (as we generally need at least 1.5 binary period cycles to trust recovery).
Our study has employed simple curve fits for binary detection as a fast and efficient methodology for rapid binary detection in large samples (e.g. as an event broker).
Future work will explore the consequences of survey design on binary parameter recovery with other detection methods.

We tested the recovery capabilities of binary periods longer than the observation baseline for a given binary target. We found that it is difficult to recover the longer periods, with most of them being underestimated by factors between 1.5 - 2.0. As seen in \cite{Davis_2024}, long-period binary signals can be misidentified with periods that are 2$\times$ or 4$\times$ lower than the actual period. In agreement with previous work, limited survey duration
can produce higher false-positive rates \citep{Vaughan2016, Barth_2018, Zhu2020, Witt_2022}. We further argue that in the case of the early years of LSST, we cannot trust the reported binary period identified by a sinusoidal curve fit detection method. Monitoring cadence, in the context of this detection method, does not significantly affect the recovery of binary period. This holds as long as the cadence is much shorter than the typical binary period and the typical damping timescale of single-quasar variability. Once 5 - 10 yrs of baseline is established, binary period estimations could begin to be trusted for most of the binary period range that we anticipate detecting with Rubin and pulsar timing. The longer a possible monitoring duration for a binary candidate, the better. The early years of LSST will benefit greatly from combining light curves from existing time-domain surveys.

As discussed above, a realistic parameter distribution anticipated for Rubin observations features more lower mass, lower luminosity quasars. 
Quasar variability amplitudes are inversely related to Eddington ratio \citep{EddingtonToVariability}. These quasars produce larger variability amplitudes. From the results compared to a fully uniform parameter grid, it will be more difficult to recover binary signals from large-amplitude quasar variability, especially if contaminated by host-galaxy light. The low-luminosity quasars do not, generally, trigger our false-positive flags with our curve-fit detection method, as noted in Figure \ref{fig:FPratesVSQuasarProperties}. From Figure \ref{fig:TPratesVSQuasarProperties}, the large variability amplitudes also result in poor true-positive recovery at long light curve durations. 

From Figure \ref{fig:SawtoothinessSpikeInvest}, we learned that we have an easier time recovering binary signals that look like the model we are fitting (smooth sines) and periodic binary signals that feature dramatic flux changes \citep[sharp pulse-like shapes from periodic accretion or self-lensing, for example][]{2018SelfLensinsDOrazio}. The work of \cite{Kelley2021} highlights that Rubin should be able to detect tens to hundreds of self-lensing binaries, we hope to be able to find them even with simple curve fits. Real binary signals may be combinations of several theoretical models, like both relativistic Doppler boosting and self-lensing \citep{2018SelfLensinsDOrazio, DOrazioCharisiBook2023}.

We found that as host-galaxy luminosity increases, the false-positive rate increases and the binary parameter recovery rates decrease. However, the overall effects are limited to changes of $< 5\%$. Host-galaxy contamination effectively ``smooths" the variability of the light curve in the ratio of the host-galaxy light to the quasar light. Smoother light curves are easier to fit, resulting in lower \redchi values for non-binary light curves. The lower \redchi values allow for more low-luminosity and low-mass quasar light curves to meet our false-positive definition as their large-amplitude variability is diluted. The suppression of variability also increases the initial recovery rates for these low-mass objects, as seen as the black points in Figure \ref{fig:ratesVcontamination}. 
The general recovery rates for all objects decrease as contamination rate increases.
Host-galaxy light will contaminate Rubin light curves until it can be characterized and subtracted.

\subsection{General Recommendations for Binary SMBH Searches with Rubin}
We recommend utilizing computationally inexpensive curve fits with caution. We further recommend the following:
\begin{itemize}
    \item Increase the monitoring duration of binary candidates to ensure accurate recovery of binary parameters and reduction in false-positive rates.
    \item Characterize and account for the host-galaxy light in order to further suppress false-positive rates for low luminosity and low mass quasars.
    \item The binary period in light curves with short monitoring duration ($<$ 5 yr) is likely to be underestimated. 
    \item Binary signals with more dramatic flux changes (i.e., sawtooth signals) are easier to identify than smoother periodic signals with the same signal amplitude, via a curve-fit detection method.
\end{itemize}


\section{Conclusion and Future Work}\label{sec:Conclusion}
We have presented our work on the creation of millions of Rubin/LSST single quasar and binary SMBH light curves and the consequences of survey design and host-galaxy light on the identification of binary SMBHs. We provide the following points as a summary of our findings:
 \begin{itemize}
     \item Compared to \cite{Davis_2024}, a more realistic parameter grid of quasar parameters and longer input binary periods results in worse, general binary parameter recovery rates. See Sections \ref{sec:ParamGridResults}.
     \item Most binary periods that are longer than the light curve baseline will not be properly recovered and will be underestimated. As in \cite{Davis_2024}, best-fit periods that are near the half-length or length of the monitoring duration should be treated with caution. See Section \ref{sec:ParamGridResults}.
     \item Longer baselines of the light curves can significantly suppress false-positive rates and improve recovery rates with a curve fit detection method. See Section \ref{sec:CadenceAndDurationlResults}.
     \item Recovery and false-positive rates remain relatively similar at a given baseline when applying different cadences. There is no best cadence for long-period binary SMBH detection with the curve-fit detection method. See Section \ref{sec:CadenceAndDurationlResults}.
     \item Generally, host-galaxy light contamination will increase false-positive rates and decrease recovery rates. The overall effective change is small, about 4-5\%, even if the host-galaxy could be nearly 10-times brighter than the quasar. See Section \ref{sec:HostGalResults}. 
 \end{itemize}

\subsection{Future Work}
The intent of this paper was to expand upon the work of \cite{Davis_2024} by introducing quasar parameter grid distributions derived from quasar population observations and stress-testing the effects of survey design and host-galaxy contamination on identifying binary SMBHs. Upcoming work includes light curves across all Rubin filters in addition to $i$-band and testing new detection methods. We also plan to apply Lomb-Scargle periodograms, Auto-Correlation Functions, and machine-learning classification \citep{Lomb, Scargle,Lombscargle, EdelsonKrolik1988,McQuillan2013ACFs, ML21}. In the future, light-curve generation could feature stronger limits on the binary SMBH parameters inferred from current gravitational-wave searches. We could also introduce more binary SMBH signals that represent more complex physical dynamics, as well as signals that evolve over the light curve lifetime. We will also explore other cadence possibilities, as well as the consequences of observation season length. The predictive methods outlined in this work serve as the foundation for application to other upcoming massive time-domain surveys, such as the Roman High Latitude Time-Domain Survey \citep[HLTDS;][]{spergel2013ROMAN, spergel2015ROMAN}.

\section*{Acknowledgements}
JRT, MCD, and LEW acknowledge support from NSF grant CAREER-1945546. MCD acknowledges support from a NSF Graduate Research Fellowship. 
MC acknowledges support by the European Union (ERC, MMMonsters, 101117624).
JCR acknowledges support from NSF award AST-2007993.
WNB acknowledges financial support from NSF grant AST-2106990.  
MCD wishes to especially thank Dani Lipman, Rachel Lee, and Hugh Sharp for their support. 

\software{ astroML \citep{astroML}, SciPy \citep{2020SciPy-NMeth}, NumPy \citep{NumPy}, LMFits \citep{lmfits}, Matplotlib \citep{Matplotlib}}


\bibliography{binary}{}

\end{document}